# Equilibration in the Aftermath of the Lunar-Forming Giant Impact

Kaveh Pahlevan* and David J. Stevenson

Division of Geological and Planetary Sciences
California Institute of Technology
MC 150-21, Pasadena, CA 91125
United States of America

**Abstract**

Simulations of the moon-forming impact suggest that most of the lunar material derives from the impactor rather than the Earth. Measurements of lunar samples, however, reveal an oxygen isotope composition that is indistinguishable from terrestrial samples, and clearly distinct from meteorites coming from Mars and Vesta. Here we explore the possibility that the silicate Earth and impactor were compositionally distinct with respect to oxygen isotopes, and that the terrestrial magma ocean and lunar-forming material underwent turbulent mixing and equilibration in the energetic aftermath of the giant impact. This mixing may arise in the molten disk epoch between the impact and lunar accretion, lasting perhaps $10^2$-$10^3$ years. The implications of this idea for the geochemistry of the Moon, the origin of water on Earth, and constraints on the giant impact are discussed.

## 1. Introduction

The Moon is generally believed to have formed from the debris ejected from the impact of a Mars-sized body onto the forming Earth [1,2]. At present, the argument in favor of such a scenario is two-fold: a single, off-center collision with a nearly formed Earth can account for the angular momentum present in the Earth-Moon system, as well as the bulk lunar iron depletion. The geochemical arguments for the giant impact, however, are less compelling than dynamical ones, because the chemical consequences of the impact have not been fully explored, with the result that compositional Earth-Moon similarities and differences -- to the extent that they trace the effects of the giant impact -- have never been satisfactorily explained.

The similarity of the inferred lunar composition to that of the silicate Earth has been taken as evidence for the derivation of the lunar material from the Earth's mantle [3]. Hydrodynamic simulations of the giant impact [4-6], however, ubiquitously derive the majority of the lunar material from the impactor. Hence, any elemental or isotopic similarity between the composition of the silicate Earth and Moon would have to be interpreted to be due to nebular mixing or planetary accretion processes. Here, we show that a compositional relationship between the silicate Earth and Moon may naturally arise in the aftermath of the impact event. The central idea is turbulent mixing of projectile and target material in the time between the giant impact and lunar accretion.

Two characteristics of oxygen make this element an ideal tracer for studying solar system formation processes. First, oxygen isotopes were distributed heterogeneously in the early

24  solar system [7] and can therefore be used to trace the sources of planetary material.
25  Second, the presence of three stable isotopes enables identification of heterogeneity that
26  occurred in a mass-independent way. Such mass-independent heterogeneity in the solar
27  system may have resulted from photochemical self-shielding of carbon monoxide in the
28  solar nebula [8,9] or its parent molecular cloud [10]. However, no petrologic process is
29  known to fractionate the isotopes in a mass-independent way, making it possible for
30  samples to precisely reflect the isotopic character of the reservoirs from which they are
31  derived. For this reason, to the extent that the sampled reservoirs are representative of
32  the composition of the parent planets, the isotopic character of the sampled planets is
33  known.
34
35  The oxygen isotopic character of the Earth and Moon are indistinguishable. Here, we
36  outline the reasons why this observation is unexpected, present a mixing model as a
37  resolution to this problem, and discuss the implications of this idea for the geochemistry
38  of the Moon, the origin of water on Earth, and constraints on the giant impact.
39
40  **2. The Problem**
41  On a three-isotope plot ($\delta^{17}O$ vs. $\delta^{18}O$), samples from a well-mixed planetary body fall on
42  a single fractionation line with slope ~0.5, reflecting mass-dependent planetary
43  fractionation processes [11]. As a consequence of the heterogeneity in the early solar
44  system, samples from distinct planetary bodies generally fall on distinct, parallel
45  fractionation lines, with the offsets from the terrestrial line measured by $\Delta^{17}O$ ($\equiv \delta^{17}O -$
46  $0.52 \times \delta^{18}O$) in parts per thousand. The isotopic signals represented by $\Delta^{17}O$ are inherited by the planets from starting materials during accretion, and are impervious to subsequent geological processes. For example, the fractionation line for samples from Mars are offset from the Earth by ~0.32 per mil [12], forming a benchmark for the length-scale and magnitude of oxygen isotope variations in planetary bodies in the inner Solar System (Figure 1). By contrast, precise measurements on lunar samples reveal that the lunar fractionation line is indistinguishable from the terrestrial line to within the analytical uncertainties [13]. Any difference between the Earth and Moon must be less than 0.005 per mil, which translates to a few percent of the difference between Earth and Mars. The Earth and Moon are isotopically indistinguishable to a very high level of precision, at least with respect to those parts of these bodies for which we have samples.

In the context of the standard giant impact scenario [14,15], the measured lunar composition places severe constraints on the composition of the impactor. The reason for this is that smooth-particle hydrodynamic (SPH) impact simulations that are successful in forming the Moon suggest that most (~80%) of the material that is injected into orbit is derived from the mantle of the impactor rather than the Earth [4-6]. Hence, even small isotopic differences between the silicate proto-Earth and impactor should have left an observable Earth-Moon difference because the impactor-derived lunar material escapes the diluting effects of mixing with the largest reservoir in the system, the Earth. Although these SPH calculations have their limitations, there is no basis to suppose that they are grossly inaccurate in this prediction. Indeed, this predominance of projectile material injected into orbit is readily understood in terms of the angular momentum of the material upon impact and the role of gravitational torques in raising the periapse of ejecta. These considerations make such an impact different from the extensively analyzed, conventional crater-forming impact events on planets. In any case, as shown below, even a 50-50 mixture would lead to difficulty in explaining the Earth-Moon similarity.

The standard explanation for the precise Earth-Moon match is that the proto-Earth and impactor formed from the same source regions. We can evaluate this possibility in the context of planet formation models and their sampling modes. Present scenarios of planet formation suggest that the process occurs in three stages: the growth of dust grains in the preplanetary disk into kilometer-sized planetesimals, the runaway growth of these planetesimals into lunar- to Mars-sized planetary embryos, and the final accretion of the planets through giant impact events [16].

During the process of runaway growth, embryos rapidly accrete all of the material in their feeding zone on ~$10^5$-$10^6$ year timescales. The feeding zones are typically only ~0.01 AU wide [17], and hence planetary embryos up to about Mars size – the size inferred for the impacting protoplanet – tend to record the oxygen isotope signature of their immediate vicinity in the disk. The formation of Earth and Venus, by contrast, requires an extended stage of planetary accumulation lasting ~$10^7$-$10^8$ years and characterized by giant impacts, a stage with a very different mode of sampling compared to that of the precursor embryos. In this last stage of growth, planetary embryos are scattered away from their places of birth. Thus, this stage of the accumulation process is accompanied by significant radial mixing. The planets that undergo giant impacts in this last stage sample material from a broad region of the protoplanetary disk. For example, the material that collects to form the Earth has significant contributions from regions interior to Mercury and exterior to Mars [18].

In this sense, the provenance of the Earth and Venus is the entirety of the inner solar system, and the composition of Earth does not reflect the composition of material at 1 AU. Hence, even if the impactor were a runaway embryo that sampled material from an annulus ~0.01 AU wide centered at 1 AU, or grew near Earth orbit at one of the Lagrange points [19], a proto-Earth-impactor difference may be expected. In particular, the Earth collects a substantial fraction (>10%) of its mass from the region of Mars, which is known to have a composition clearly distinct from the Earth. If the impactor accreted a few percent more or less material from the region of Mars than did the Earth, an Earth-Moon difference would have been observed. Stochastic, large-scale sampling, together with large-scale heterogeneity evinced by the Earth-Mars difference, strongly argues for heterogeneity among planetary embryos.

To quantify this argument, we can take the accretion simulations of Chambers [18] and forward model the composition of the impactors onto the forming planets. In this procedure, we take the starting embryos in the simulations and assign values of $\Delta^{17}O$ based on heliocentric distance. We then follow the accretion through collisions and trace the provenance of the planets and the impactors that contribute to their formation. In this way, we can make statements about the compositional scatter of the embryos that participated in giant impacts with the Earth during its formation. We do not know what the isotopic gradients in the inner solar system looked like. However, we do know that there was heterogeneity on the scale of planets, and any proposed initial profile must be consistent with this constraint. The simplest assumption is that there was a linear gradient with heliocentric distance in the inner planet region:

$$\Delta^{17}O(r) = c_1 \times r + c_2 \qquad (1)$$

One advantage with this assumption is that there are only two free parameters, both of which can be calibrated so that, at the end of the accretion simulation, the third planet from the Sun has the composition of Earth ($\Delta^{17}O = 0‰$) while the next planet out has the composition of Mars ($\Delta^{17}O = +0.32‰$). For this purpose, we choose a simulation from Chambers [18] that yields four terrestrial planets at its conclusion. The result of this calculation is a histogram of the compositional scatter of impactors onto the forming planets, calibrated to the Earth-Mars difference (Figure 2). In order to overcome small-number statistics, we include the compositional scatter of impactors onto all four final planets. The main conclusion from this approach is that the scatter among the impactors onto the planets is comparable to the observed differences between the planets. In particular, none of the planetary impactors in this simulation has an isotopic composition similar enough to the final planet to yield the Earth-Moon system.

There are several questions raised by this approach that must be addressed. First, since the oxygen isotope composition of the gas (in CO and $H_2O$) may be different from that of the solids (in silicates and oxides), assigning an isotopic composition to a heliocentric distance is only meaningful if the solids are isotopically disconnected, i.e. no longer equilibrating with the gas, and one is referring only to the solids. Second, the oxygen isotope composition of Vesta ($\Delta^{17}O$ = -0.22‰) [20], at its present heliocentric distance of 2.4 AU, is inconsistent with a monotonic radial variation in the early solar system. However, a recent model for the formation of differentiated asteroids suggests that Vesta formed in the terrestrial planet region and was later scattered into the asteroid belt [21]. Since the orbits of the less massive minor planets can be relatively easily shifted through gravitational scattering, the composition of Vesta is not a strong argument against an initial monotonic gradient in the inner planet region. In any case, both because of its mass and its proximity, we believe that the composition of Mars is a better guide to the gradients that prevailed in the terrestrial planet region during Earth formation.

To summarize, the present composition of the Earth reveals the average composition of contributing impactors, and approximates the composition of the inner solar system as a whole. In the standard picture of planet formation, Mars is a remnant of a population of planetary embryos that collided to form the Earth and Venus. The precise match between the composition of the Earth and Moon is difficult to reconcile with the heterogeneity observed between the terrestrial planets and present scenarios of accretion. The question we now address is whether post-impact mixing processes could have homogenized the Earth-Moon system, reducing any pre-existing isotopic heterogeneity to undetectable levels.

**3. Model**

162  Immediately after the giant impact, the Earth-Moon system is largely molten and partially

163  vaporized [15]. The silicate Earth melts and forms a deep magma ocean, the proto-lunar

164  material forms a circumterrestrial magma disk, and a common silicate-vapor atmosphere

165  links the disk to the planet (Figure 3). The terrestrial magma ocean and the proto-lunar

166  magma disk, as well as the enveloping silicate vapor atmosphere, are vigorously

167  convective due to the high heat flow needed to match radiative losses from the

168  photosphere of the planet and disk. Here we show that under such conditions, mixing

169  and equilibration of the Earth's mantle with the proto-lunar disk is possible, and that the

170  terrestrial and lunar material may have approached diffusive equilibrium with respect to

171  their isotopic composition.

172

173  The gravitational energy released during the giant impact is large. The timescale to

174  eliminate this heat – and hence to condense the silicate vapor atmosphere – is determined

175  by radiative cooling:

176

177  $$\tau_{cool} = \frac{GM_E m_I}{\sigma T_e^4 4\pi R_E^3} \qquad (2)$$

178

179  An effective photospheric temperature of ~2,000 K [22] and an impactor mass ~0.1 $M_E$

180  [5] yields a cooling timescale for the Earth of ~$3 \times 10^3$ years. The corresponding timescale

181  for the proto-lunar disk will be shorter than this estimate because (1) the surface area of

182  the impact-generated disks are typically greater than that of the Earth and (2) only a

183  fraction of the energy of impact is partitioned to the orbiting material. Hence, the cooling

184  timescale for the lunar material will be somewhat shorter, perhaps $10^2$-$10^3$ years. In this respect, it is important to emphasize that the thermal energy deposited as heat in the orbiting material – although enough to partially vaporize it – is small relative to the latent gravitational energy that may be gradually released if the disk viscously evolves [22]. We will refer back to this point later.

The extended disk atmosphere is contiguous with the silicate vapor atmosphere of Earth and can therefore act as an exchange medium, communicating isotopic signals between the two liquid reservoirs with which it is in contact. Although constituting only a fraction (maybe 10%) of the lunar disk by mass, and an even smaller fraction of the Earth, the vapor atmosphere may process enough material to equilibrate the massive liquid reservoirs through continuous exchange. Since the silicate vapor atmosphere exists for a finite interval, any diffusive mixing must occur within such a timescale. We adopt the conservative point of view that there is negligible direct transport between the liquid in the disk and the terrestrial magma ocean, consistent with [22]. Below, we attempt to estimate the rates of various mixing processes, and to show that they are fast in comparison with the cooling timescale.

3.1. Convection within the Earth, disk, and common atmosphere

The energy released during the giant impact will heat both the Earth and the projectile by many thousands of degrees [6]. Large-scale motions will be turbulent and even small-scale turbulence will be possible because the silicate melt will be hot and de-polymerized. The heat flux from the system will be limited by radiation from an atmosphere with a photospheric temperature of ~2,000 K [22]. A crude estimate for the convective

velocities required to accommodate these heat fluxes can be derived from mixing length theory (MLT):

$$V_{conv} = \left(\frac{FL}{\rho H}\right)^{1/3} \qquad (3)$$

Here F is the convective flux (equal to the radiative flux in steady-state), $\rho$ is the density of the convecting medium, H is the temperature scale height of the system, and L is the "mixing length" which represents the length scale associated with the largest eddies. In stellar convection, where MLT has found common application, the mixing length is often taken to be some significant fraction of the scale height of the system. Hence, L/H, and especially $(L/H)^{1/3}$ will be of order unity. We take L/H = 0.1. Using this approach, we can estimate the convective velocity knowing only the temperature at the disk photosphere and the density of the convecting liquid or vapor. For a silicate liquid with $\rho \sim 3 \times 10^3$ kg/m$^3$, the convective velocity is ~3 m/s whereas a silicate vapor with density $\rho \sim 3$ kg/m$^3$ [22] carrying the same heat flux requires a convective velocity of ~30 m/s. The flux decreases to zero at the disk mid-plane, but this will not affect these estimates for the largest-scale eddies. Rotation may inhibit the length-scales somewhat, through the Coriolis effect, but if one chose L so that $v/\Omega L \sim 1$ where $\Omega$ is the Keplerian rotation, then this predicts a smaller velocity by a factor of only a few.

The corresponding turnover timescales ($\equiv L/v$) are: a *week* for the Earth's mantle, and several *hours* in the magma disk and vapor atmosphere. Turnover timescales turn out to be much faster than the timescales for other processes, discussed below. Vertical mixing *within* the Earth, disk, and vapor atmosphere is fast compared to the cooling timescale, and is unlikely to be the rate-limiting step in Earth-Moon equilibration. Such rapid turnover makes it possible for material in each reservoir to be tapped for exchange through continuous exposure to liquid/vapor interfaces.

## 3.2. Liquid/Vapor Exchange

In this section, we attempt to estimate the timescale for the evaporative exchange of atoms between the liquid and vapor phases, both at the Earth-atmosphere interface, and the disk-atmosphere interface. This timescale is important because the vapor phase is the main carrier of isotopic signals between Earth and disk, but represents only a fraction, (maybe 20%) of the disk mass, and an even smaller fraction of an Earth mass. Hence, for the signals to be successfully communicated between the liquid reservoirs, there must be continuous, efficient liquid-vapor exchange.

In thermal equilibrium and with a slow change in mass fraction of the co-existing phases, the flux of atoms across the phase boundary is nearly the same in either direction. Evaporative exchange from the gaseous phase to the liquid can be estimated using the kinetic theory of gases. Aside from a constant numerical factor, the flux of gas molecules incident on the liquid surface is simply the product of the number density and the average thermal velocity. To get the exchange rate, this kinetic rate must be multiplied by a condensation coefficient, which represents the fraction of molecules impinging on the liquid surface that enter the liquid phase. Assuming ideal gas behavior for the vapor, the timescale to exchange the mass of the magma disk with the atmosphere is given by:

$$\tau_{ex} = \sigma V_t / P\alpha_c \tag{4}$$

where $\sigma$ is the surface mass density of the magma, $V_t$ is the molecular thermal velocity, P is the vapor pressure, and $\alpha_c$ is the condensation coefficient. For the most abundant silicate mineral in planets ($Mg_2SiO_4$, forsterite), its numerical value is ~0.1 [23]. For a ~2 lunar-mass disk extending out to ~5 Earth radii, $\sigma$ ~ 5 x $10^7$ kg/m$^2$. The vapor pressure of silicates at temperatures of ~3,000 K is ~10-100 bars [22]. Finally, for a forsterite vapor composed of MgO, SiO, and $O_2$ [14], the mean molecular weight is ~40 amu, which yields a thermal velocity of ~1 km/s. These parameters, which apply to the Earth-atmosphere interface as well as the disk-atmosphere interface, suggest that the exchange of several lunar masses of material across the phase boundary requires a timescale of the order of a week.

However, rapid exchange across a liquid-vapor interface does not guarantee rapid equilibration of the phases. The reason is that in the absence of diffusion or fluid motions within each phase, the *same* atoms are exchanged across the phase boundary. In this respect, it is important that the exchange is occurring in a two-phase medium (the liquid is interspersed with vapor bubbles, and the vapor with liquid droplets) that enhances the surface area for evaporative exchange compared to a smooth liquid/vapor interface [22].

275  Although the disk may undergo phase separation, allowing the liquid to settle [24], the
276  viscous evolution of the disk liberates enough energy to vaporize the disk [15]. Indeed,
277  as emphasized by [22], the disk will be in a thermally regulated state such that the
278  viscous self-heating balances the radiative cooling. Hence, in addition to the vapor that
279  condenses at high levels in the atmosphere, generating clouds and droplets, part of the
280  liquid magma vaporizes, generating bubbles. Since the timescale for the rainout of
281  droplets may be as short as a week [24], advective transport between liquid and vapor
282  may be determined by the rate at which the vapor condenses into droplets via radiative
283  cooling. The timescale to condense the mass of the vapor atmosphere, and hence to
284  advect the vapor composition to the liquid disk is:

285

286 $$\tau_{rain} = L\sigma_v / F \qquad (5)$$

287

288  where L is the latent heat of condensation for silicates, $\sigma_v$ is the surface density of the
289  vapor atmosphere, and F is the radiative flux. For L ~ $10^7$ J/kg, $\sigma_v$ ~ $10^7$ kg/m$^2$
290  corresponding to a disk composed of ~20% vapor, and F determined by an effective
291  radiating temperature of ~2,000 K [22], this timescale turns out to be of order one year.
292  As mentioned earlier, the latent gravitational energy gradually liberated by the disk
293  viscously exceeds its thermal energy, allowing many rainout cycles to take place before
294  the disk cools and condenses. Thus, the continuing process of phase separation amounts
295  to a rapid, vertical advective transport between the liquid and the vapor, and may be
296  instrumental in equilibrating the two phases.

297

In summary, the liquid/vapor exchange across the phase boundary is fast – it occurs on a timescale of a week – but this does not guarantee phase equilibration. Liquid/vapor exchange is facilitated enormously by the fact that the process is taking place in a two-phase medium. As an example, air/sea exchange on the present Earth occurs much faster than would be expected due to the presence of bubble plumes near the ocean surface [25]. An analogous process in the lunar disk might enable equilibration of the liquid and vapor on a timescale of years. Hence, liquid/vapor exchange is unlikely to be the rate-limiting step in the equilibration of the Earth and Moon.

3.3. Exchange from the Earth to Disk

Although the common atmosphere surrounding the Earth and disk is continuous, it nevertheless can be dynamically distinguished into two distinct regimes. The silicate vapor atmosphere surrounding the post-impact Earth is mainly supported against gravity by pressure gradients, whereas the disk atmosphere is supported by pressure gradients in the vertical direction, but mainly supported by nearly Keplerian rotation in the radial direction. Since the Earth will be rotating significantly slower than breakup velocity, there will be a velocity shear of several kilometers per second that separates these two dynamical regimes in the common vapor atmosphere.

The shear instability that results from this velocity difference is likely to facilitate a constant exchange across this dynamical interface. However, the gas density in this region may be lower than elsewhere because the atmospheric scale height is much smaller than the planetary radius. For the present, we make the assumption that is most

321 favorable to the mixing hypothesis, namely, that the exchange across this region is faster
322 than elsewhere.
323
324 3.4. Radial Mixing within the Disk
325 We have argued above that the processes necessary for Earth-Moon equilibration through
326 exchange with a vapor atmosphere: convection, liquid/vapor exchange, and exchange
327 across a dynamical interface, are possibly fast in comparison with the cooling timescale.
328 If this is indeed the case, then the rate-limiting step for Earth-Moon equilibration will be
329 radial mixing through the proto-lunar disk. The reason for this is that the Moon forms
330 from the outer-most disk material [26], and so it is important to quantify the extent to
331 which the inner regions of the disk can communicate their terrestrial isotopic signals to
332 the regions from which the Moon forms.
333
334 We assume that the liquid disk and its vapor atmosphere are turbulent, and that the fluid
335 motions can be characterized by an eddy diffusivity [27]. We can then write Fick's Law:
336
$$\vec{J} = -\rho f_v D \vec{\nabla} c \tag{6}$$
338
339 where J [kg m$^{-2}$ s$^{-1}$] is the mass flux, $\rho$ is the density of the two-phase fluid, $f_v$ is the vapor
340 fraction, D is the turbulent diffusivity of the vapor, and c is the mass fraction of a passive
341 tracer. Here, we have assumed that the liquid is stationary while the vapor is diffusive
342 because the vapor will be undergoing more vigorous convective motions than the liquid
343 (section 3.1). This equation can nevertheless describe changes in the composition of the liquid if there is rapid vertical equilibration of the liquid and vapor phases, i.e. the signals carried by the vapor are communicated to the liquid (section 3.2). We take the divergence of equation (6) and combine it with the continuity equation to get:

$$\frac{\partial(\rho c)}{\partial t} = \vec{\nabla} \cdot (\rho f_v D \vec{\nabla} c) \qquad (7)$$

For simplicity, we only consider a static disk, so that the density will be time-independent and can be taken out of the time-derivative. This assumption is valid as long as we consider timescales shorter than the timescale for the evolution of the disk, which may be $10^2$-$10^3$ years [22]. Finally, because very little mass lies above the radiative photosphere, most of the disk will participate in the convective fluid motions. Hence, to a good approximation, we can integrate equation (7) in the direction perpendicular to the disk plane to yield:

$$\sigma \frac{\partial c}{\partial t} = \vec{\nabla} \cdot (\sigma f_v D \vec{\nabla} c) \qquad (8)$$

where $\sigma$ [kg m$^{-2}$] is the total surface density. We can solve this equation in cylindrical coordinates with two boundary conditions. (1) The composition at the inner boundary of the disk is taken to be equal to the terrestrial composition, and constant with time, i.e. $c(r = R_E, t) = 0$. Even though c represents a mass fraction, we can shift the terrestrial composition to zero because the diffusion equation is linear. The time-independence of this boundary condition reflects the circumstance that convection within the Earth

(section 3.1), liquid-vapor exchange of terrestrial silicates (section 3.2) and vapor exchange at the Earth-disk interface (section 3.3) are efficient processes, enabling the entire terrestrial magma ocean to be tapped. This condition is satisfied to good approximation as long as the mass of the post-impact Earth that participates in the equilibration is much greater than the mass of the lunar disk. (2) There is zero net-flux at the outer boundary of the disk, i.e. $\partial c/\partial r(r=R_{out},t)=0$. This condition stems from the assumption that all fluid parcels that reach the outer disk boundary reflect back and continue to participate in the turbulent motions. In effect, we neglect condensation and moonlet formation from the outer edge of the disk, possibly a valid assumption as long as we consider timescales shorter than the cooling time.

To solve this equation, we need to know the vapor fraction, $f_v$, and also something about the mass distribution, $\sigma(r)$. One possible approach uses the results from the impact simulations, which yield disk vapor fractions of ~20% [6] and show that $\sigma(r)$ decreases roughly as ~1/r (pers. comm., Robin Canup). To estimate the vigor of turbulent mixing, we parameterize the diffusivity in terms of the alpha model and present our results in terms of the alpha parameter. Later, we use mixing length theory to estimate a value for alpha that may be realistic. The diffusivity can be written as:

$$D = \alpha c_s H \tag{9}$$

where $c_s$ is the gas sound speed, which, for a temperature of ~2,500 K and a mean molecular weight of ~40 amu for silicate vapor is ~1 km/s. H is the pressure scale height

389  of the disk atmosphere, which corresponds to ~$10^3$ km in the inner regions of the disk and
390  increases as ~$r^{3/2}$ in our model, as expected for a roughly isothermal disk.  $\alpha$ is a
391  dimensionless number that parameterizes the vigor of the turbulent motions.  This
392  parameter is often introduced in the context of "viscous" disk evolution [28, 29] but it is
393  important to understand that we are here using it for mixing, not net mass transfer.  Both
394  processes may occur, but the net mass transfer could be zero and our model still work.  In
395  fact, the ability of turbulence to redistribute angular momentum is questionable [30].
396
397  **4. Results**
398  In the previous section, we have argued that the rate-limiting step for Earth-Moon
399  equilibration is radial mixing through the proto-lunar disk.  Here, we present results of
400  calculations of radial mixing subject to the assumptions described above.  Figure 4 shows
401  snapshots of the disk composition at various times.  Initially, exchange with the Earth
402  causes the inner regions of the disk to become contaminated with terrestrial composition.
403  As the diffusion proceeds, the composition of the disk becomes progressively more
404  Earth-like.  Turbulent mixing may proceed for a time period comparable to the cooling
405  timescale, after which diffusive exchange between the Earth and disk halts due to the
406  condensation of the vapor into the liquid phase.  The mean composition of the Moon is
407  then obtained by integrating the composition of the outer half of the disk.
408
409  The longer the turbulent diffusion proceeds, the greater is the dilution of the Earth-Moon
410  difference from post-impact levels.  We can define an equilibration time as the timescale
411  to reduce the Earth-lunar disk difference by a factor of 10-100.  (From Figure 2, the

412   majority of impactors require dilution of such magnitude to yield the measured lunar
413   composition). Figure 5 displays the quantitative trend of how the vigor of turbulence
414   reduces the equilibration time. The main result from these calculations is that for
415   timescales of $10^2$-$10^3$ years, efficient mixing between the Earth and the lunar-forming
416   material requires alpha values of $10^{-3}$-$10^{-4}$.

417

418   What is a realistic value for alpha? To estimate the vigor of turbulence in the lunar disk,
419   we must have some idea of the physics that underlies the instability that leads to
420   turbulence. If we assume that thermal convection is the primary source of turbulence, we
421   can calculate the diffusivity as the product of the convective velocity, derived from
422   mixing length theory, and the length scale associated with the convection:

423

424 $$D = V_c L \qquad (10)$$

425

426   Here, L is the 'mixing length', which represents the size of the largest convective eddies,
427   and is typically taken to be some significant fraction of the scale height of the system.
428   Here we take L/H = 0.1. For such a diffusivity, we get an alpha value of ~$3\times10^{-4}$
429   suggesting that efficient turbulent mixing across the extent of the lunar disk is possible.
430   This simple estimate ignores the effects of rotation, which may inhibit radial diffusion,
431   and should be taken as an upper limit. As noted previously, modestly smaller length
432   scales (i.e. L ~ 10 km) may not change the velocity estimate much but in equation (10)
433   such a smaller eddy size might prevent our mechanism from working.

434

## 5. Discussion

These calculations suggest that extensive mixing between the Earth and the lunar disk is possible. However, there are unresolved questions that prevent a more definitive conclusion. For example, since the Moon forms from the outermost disk material, it may not fully participate in the diffusion occurring between the inner regions and the Earth. In particular, the outermost regions of the disk cool faster, perhaps freezing and cutting off a fraction of the proto-lunar material from isotopic exchange. Furthermore, it is not yet clear whether efficient exchange of material between the terrestrial and disk atmospheres occurs. Despite these uncertainties, it is possible that the Moon will form with a significantly more Earth-like oxygen isotope signature than the impactor that triggered its formation.

Current scenarios of planet formation suggest that the Earth and the impactor are unlikely to have had the same composition to within the analytical measurements. Here, we have explored the possibility that such formation theories are correct, and that the Earth-Moon system equilibrated in terms of oxygen isotopes in the aftermath of the giant impact. However, it is also possible that current scenarios of late-stage planet formation are incomplete, and that the predictions they make regarding radial mixing are incorrect. In particular, there are processes, such as non-accretionary collisions, dynamical friction with a sea of small bodies, and interaction with a small amount of residual nebular gas, that are not fully incorporated in present dynamical simulations, and that may be important in determining the final provenance of the terrestrial planets.

458  This raises the possibility that Mars is anomalous, that is, the Earth-Mars difference is not
459  representative of the scatter between the embryos that collided to form the Earth. In the
460  present formation scenarios, the source regions of the Earth and Mars overlap, and the
461  Earth accretes a significant amount of material from the Mars region, with compositions
462  that presumably match that of Mars. However, the same dynamical simulations are
463  unable to produce planets with masses as low as Mars or Mercury [18]. It has been
464  recognized that the depletion in mass from the region of Mars may be the tail end of the
465  depletion of the asteroid belt, and may not be a feature that arises from late-stage
466  accretion. In this regard, it is worthwhile to attempt to track the fate of material
467  originating in the region of Mars. In particular, while >10% of the planetary embryos in
468  the asteroid belt that are removed by mean-motion resonances collide with the terrestrial
469  planets, those embryos that are removed by the secular resonance at 2.1 AU are
470  dynamically excited so rapidly that they almost always collide with the Sun [31]. It is
471  important to continue to consider the fate of Mars-like material in order to clarify the
472  meaning of oxygen isotopes for planet formation.

473

474  Although oxygen isotope measurements are at present not available for Venus and
475  Mercury, the dynamical scenarios predict that rocks from Venus, which also sampled a
476  wide region of the inner solar system, will fall on a fractionation line close to, but not
477  identical with, that of the Earth-Moon system. Similarly, the scenario outlined here,
478  whereby it is assumed that the innermost solar system exhibits heterogeneity in oxygen
479  isotope abundances, suggests that Mercury, which was the product of a few runaway
480  embryos, can be expected to have an oxygen isotope anomaly comparable in magnitude to that of Mars. The identification of meteorites or the successful completion of sample return missions from these planets may one day enable us to test these predictions.

## 6. Implications

6.1 Lunar Geochemistry

What are the implications of the proposed model for the geochemistry of the Moon? The turbulent mixing and equilibration that is invoked to explain the Earth-Moon similarity in oxygen isotopes is not restricted to the element oxygen, but may include other tracers of terrestrial mantle composition, for example, silicon [32]. If the chemical composition of the liquid and vapor were the same, then isotopic homogeneity through turbulent mixing between the Earth and Moon would necessarily imply chemical homogeneity as well. However, equilibrium thermodynamics dictates elemental fractionation between the liquid and vapor. For example, it has been experimentally determined [33] that fayalite ($Fe_2SiO_4$) has a higher vapor pressure than forsterite ($Mg_2SiO_4$). For this reason, it is likely that the silicate vapor will have a higher Fe/Mg ratio than the silicate liquid with which it is in contact. This can cause a compositional difference between silicate Earth and Moon even in a closed system where no vapor escapes to infinity. Although a detailed treatment of the chemical consequences of equilibration is beyond the scope of this paper, we merely note that melt-vapor equilibrium includes isotopic, as well as elemental fractionation. Hence, major-element chemical differences between the silicate Earth and Moon [34] will, in this model, be accompanied by mass-dependent isotopic differences. Determining the expected magnitude of such equilibrium isotopic fractionation will be the topic of future research.

## 6.2. Origin of Water on Earth

Among the chief differences between Earth and Moon is the stark depletion of volatile elements on the Moon, including water. The model we have put forward suggests that the Earth should have transmitted a volatile-element signal to the lunar material. However, the lunar rocks are strongly depleted in volatile elements relative to the terrestrial mantle [3]. The resolution to this dilemma may be the consideration of open-system processes such as hydrodynamic escape. Hydrogen in the lunar disk will be unbound [14], and it is well known that a hydrodynamic wind of light elements is capable of entraining heavy elements (i.e. Na, K) in the outflow that would not escape of their own accord [35]. Whether or not a particular element significantly escapes depends on its volatility, abundance, and atomic mass. To determine whether the proposed mixing scenario can be reconciled with the lunar depletion of volatile elements, it will be necessary to take such considerations into account. Here, we merely note that if the scenario that we have outlined is correct, the lunar depletion of volatile elements may require the Earth to transport hydrous materials to the lunar disk to drive an outflow, a constraint that requires the Earth to have accreted significant amounts of water before the moon-forming giant impact.

## 6.3 Constraints on the Giant Impact

Among the most striking differences between the Earth and Moon is the bulk lunar depletion of iron (by ~3x from cosmic composition). The present explanation for this depletion is the differentiation of the impactor prior to the giant impact, and the preferential injection of impactor mantle into circumterrestrial orbit. However, in the scenario that we have outlined here, turbulent mixing and equilibration with the terrestrial mantle may remove iron from the orbiting disk, obviating the necessity for the giant impact to directly account for this fundamental Earth-Moon difference. Although not demonstrated here, it may be possible to relax this constraint on giant impacts capable of producing an iron-depleted Moon. At the same time, the necessity of equilibrating oxygen isotopes may require an additional constraint. Current impacts place roughly an equal amount of material interior and exterior to the classical Roche limit [6] defined by the lunar bulk density at 2.9 Earth radii. Inside this radius, tidal forces prevent the disk material from aggregating. Although a small amount of vaporization will greatly reduce the density and move the Roche radius outward, outside of the classical Roche radius, pure melts/solids may accrete when they collide. For this reason, a more massive, Roche-interior initial disk may be a more conducive starting condition for equilibration.

## 7. Conclusions

Dynamical simulations of terrestrial planet formation suggest that the Earth and the moon-forming impactor did not have identical source regions. The heterogeneity implied by the oxygen isotope difference between Earth and Mars suggests that the impactor may have been isotopically distinct. The model we have put forward suggests that in the aftermath of the giant impact, the proto-Earth and the proto-lunar disk may have approached diffusive equilibrium, reducing any pre-existing differences in oxygen isotope composition, and perhaps eliminating any primary heterogeneity in the Earth-Moon system. This model has testable consequences for the geochemistry of the Earth and Moon. Unraveling the oxygen isotope story in the inner solar system, for example, by sampling Venus and Mercury, may help to resolve the long-standing problem of the provenance of the terrestrial planets. The possibility explored here is a first step towards understanding the meaning of oxygen isotopes for planet formation.

**Acknowledgements**

We would like to thank Alessandro Morbidelli for bringing this problem to our attention, John Chambers for kindly sharing detailed simulation results, Robin Canup, John Eiler, and Colette Salyk for insightful comments, and Herbert Palme and an anonymous reviewer for helpful suggestions on the manuscript. We would like to dedicate this paper to the memory of Ted Ringwood, an early advocate of a terrestrial origin of lunar matter.

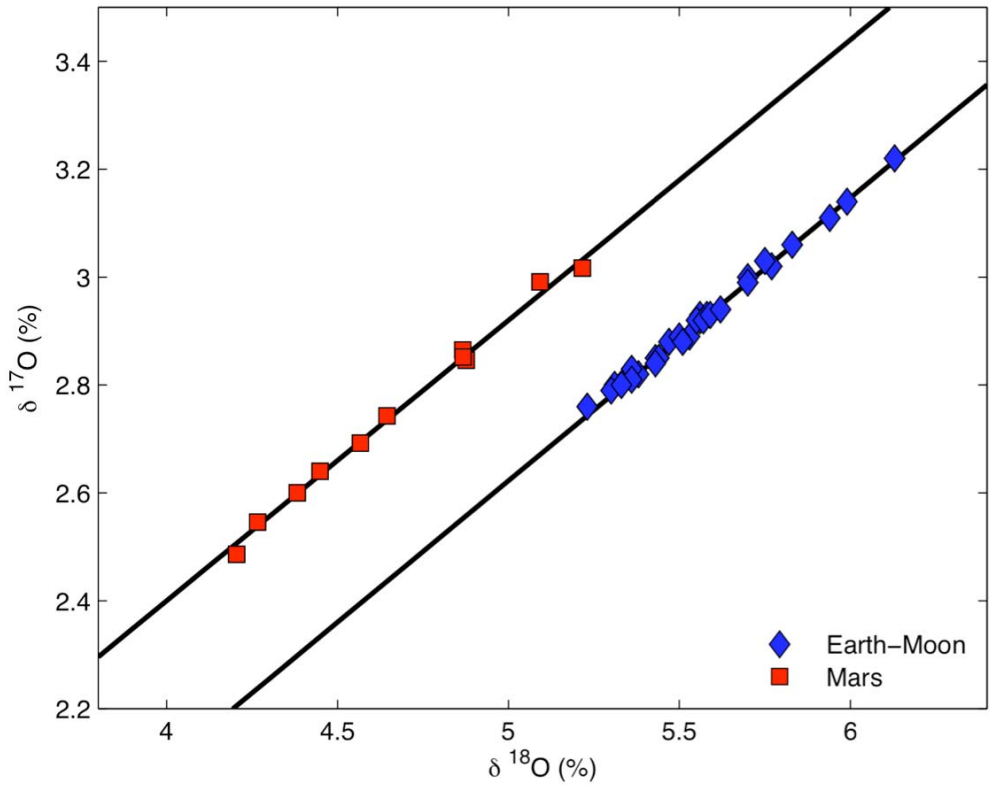

**Figure 1.** Oxygen isotopes for the Earth-Moon system and Mars. Samples from Mars are offset from the Earth by 0.32 per mil, while lunar data are indistinguishable from the Earth at the level of 0.005 per mil. Data from [12,13]

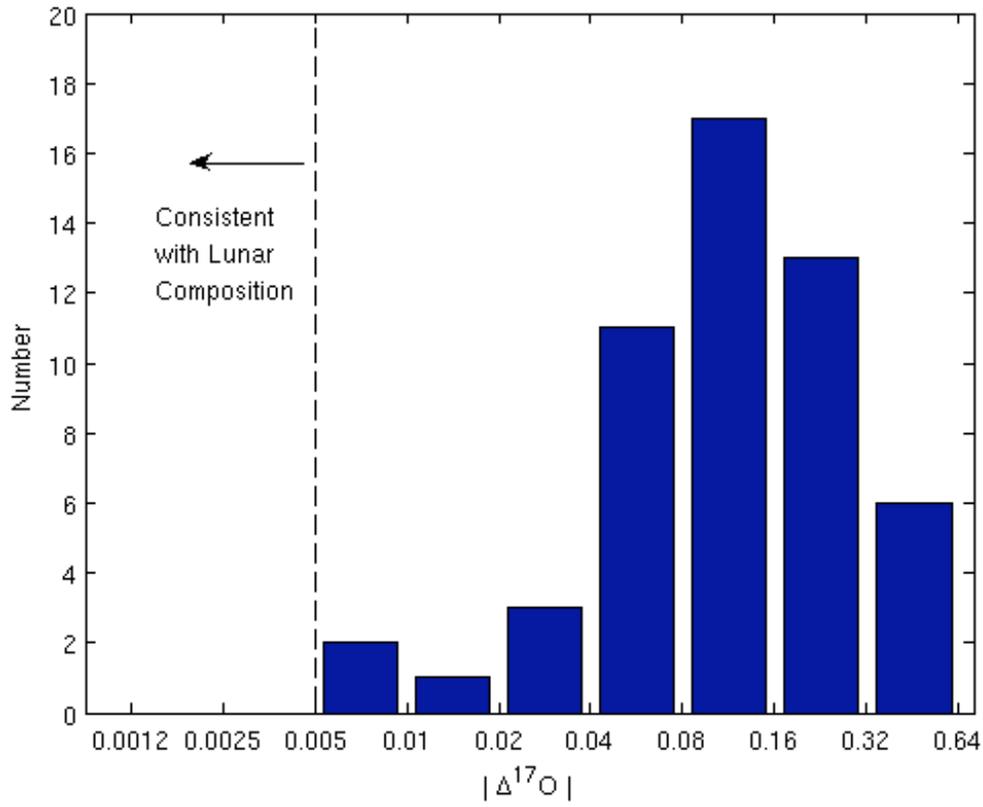

**Figure 2.** Histogram of the composition of impactors onto the planets in Chambers' [18] simulation 21, calibrated to the Earth-Mars difference. The mean deviation of the impactors' compositions from the final planets is $<| \Delta^{17}O_{imp} - \Delta^{17}O_{planet} |> = +0.15‰$. Compositional scatter among incoming impactors is comparable to the differences observed between the planets. None of the impactors in this simulation had a composition similar enough to the target planet to yield the Earth-Moon system. A linear gradient with heliocentric distance is assumed.

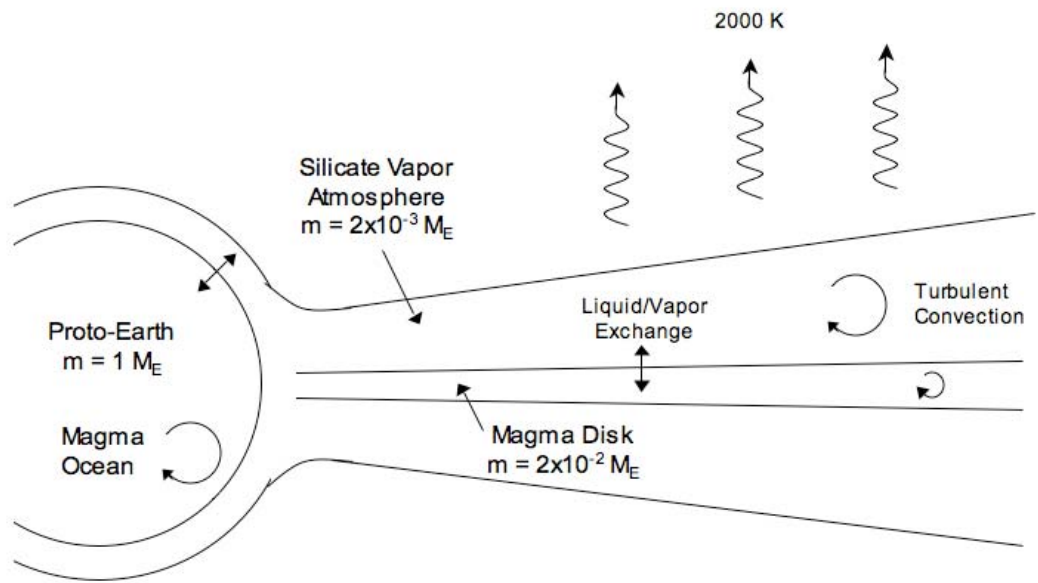

**Figure 3.** Schematic of Earth and proto-lunar disk immediately after the giant impact. High radiative heat loss guarantees convection in the silicate Earth, disk, and atmosphere. Liquid/vapor exchange with a common silicate vapor atmosphere makes it possible for the two massive liquid reservoirs to equilibrate. Convection within the Earth allows the entire terrestrial silicate reservoir to be tapped.

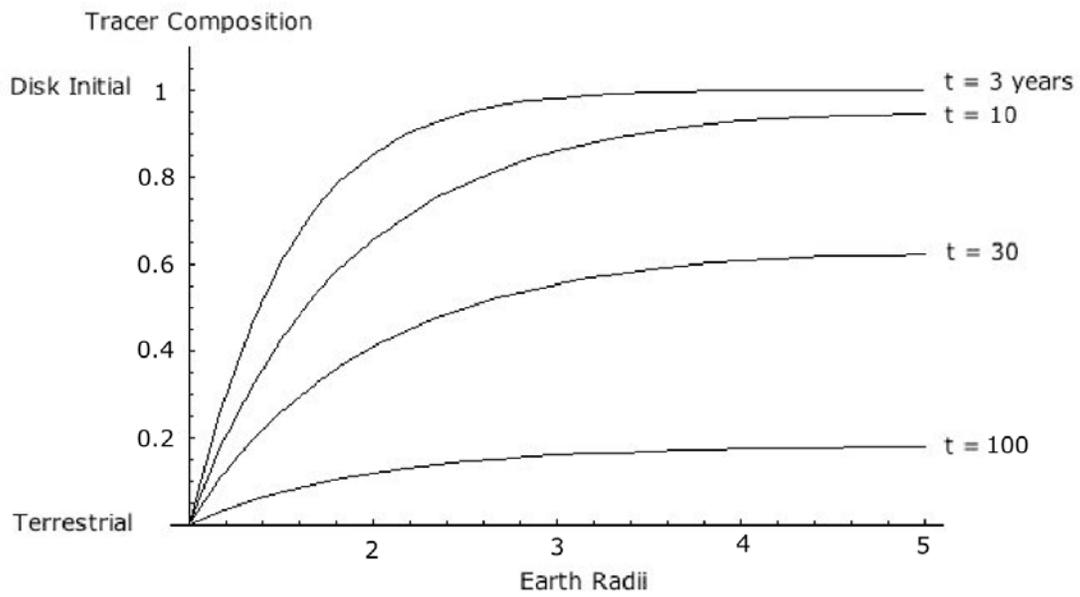

**Figure 4.** Snapshots of the solution to the diffusion equation, with the value of the tracer composition plotted against the cylindrical radius. The Earth's tracer composition is defined as zero and is unchanged by mixing; the initial composition of the disk is unity. Such mixing applies to passive tracers that readily partition between the liquid and vapor.

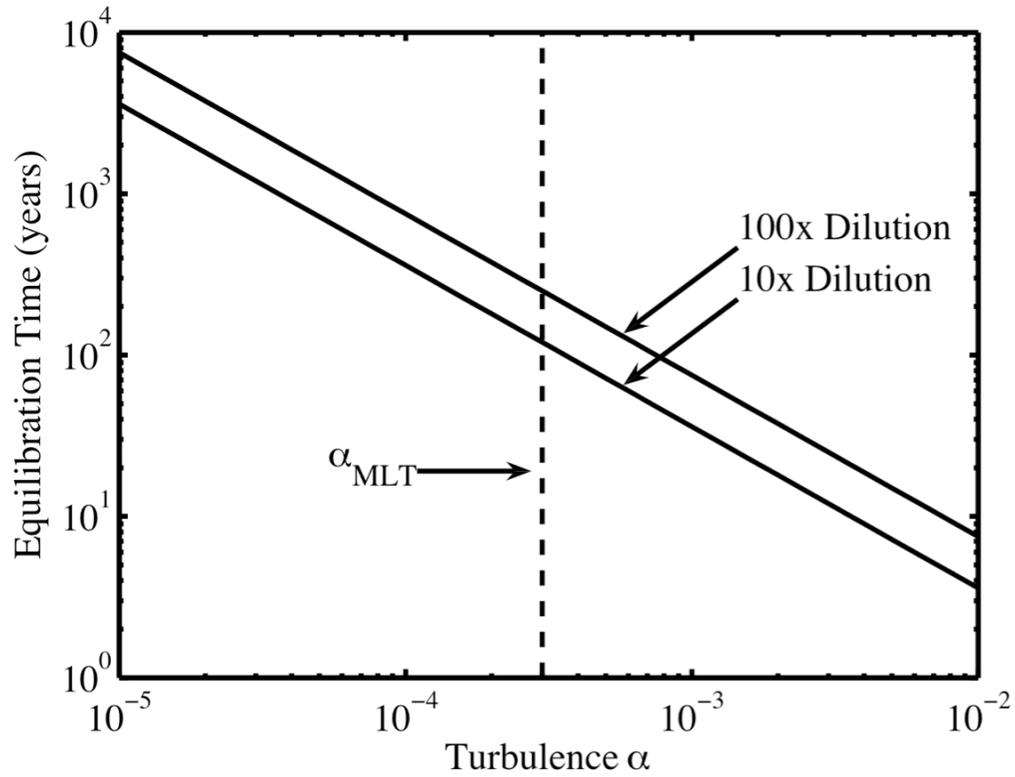

**Figure 5.** Timescale to dilute the Earth-Moon isotopic difference as a function of the diffusive turbulence alpha parameter. The lunar composition is obtained by integrating the composition of the outer half of the disk. Mixing length theory gives $\alpha \sim 3 \times 10^{-4}$ corresponding to 100x dilution in ~250 years. The equations for the two lines are $T(\text{years}) = 0.75/\alpha$ and $T(\text{years}) = 0.36/\alpha$.